\documentclass[prb,reprint,superscriptaddress,twocolumn]{revtex4-2}
\usepackage{amsmath,amssymb,bm,bbm,braket,mathrsfs,mathtools}
\usepackage{multirow}
\usepackage{color}
\usepackage{url}
\usepackage{hyperref}
\hypersetup{
	setpagesize=false,
	bookmarksnumbered=true,%
		bookmarksopen=true,%
	colorlinks=true,%
	linkcolor=magenta,
	citecolor=magenta,
}

\DeclareMathOperator{\tr}{Tr}

\DeclareMathOperator{\sgn}{sgn}
\DeclareMathOperator{\pf}{Pf}

\begin{document}
\preprint{APS/123-QED}

\title{Representation-protected topology of spin-singlet $s$-wave superconductors}
\author {Shingo Kobayashi}
\affiliation{RIKEN Center for Emergent Matter Science, Wako, Saitama, 351-0198, Japan}
\author{Akira Furusaki}
\affiliation{RIKEN Center for Emergent Matter Science, Wako, Saitama, 351-0198, Japan}
\affiliation{Condensed Matter Theory Laboratory, RIKEN, Wako, Saitama, 351-0198, Japan}
\date{\today}

\begin{abstract}
We show that spin-singlet $s$-wave multi-band superconductors have a topological phase protected by rotation symmetry and time-reversal symmetry without spin-orbit coupling in two and three dimensions.
This topological phase, an example of a representation-protected topological phase, has a $\mathbb{Z}_2$ topological index and is stable as long as the bands at the Fermi energy are formed by a doublet of orbital states with finite angular momenta.
In the limit of weak superconducting pair potential, the $\mathbb{Z}_2$ index gives a Fermi-surface formula and is related to the winding number of three-dimensional strong topological superconductors of class CI.
We present a model of a topological $s_{\pm}$-wave superconductor that has gapless surface states with a quadratic dispersion and suggest a connection with iron-based superconductors.   
\end{abstract}
\maketitle

\paragraph*{Introduction.} 
The search for topological superconductors is one of the central challenges in the field of superconductivity~\cite{Hasan-Kane2010,Qi-Zhang2011,Tanaka12,SatoFujimoto16,Sato17}, 
and much effort has been devoted to the study of $p$-wave superconductors with spin-orbit coupling~\cite{Read00,Kitaev01,Sato03,SFT09,Sau10,Lutchyn10,Oreg10,Jason10} 
and superconducting topological materials~\cite{Sato2010,Fu2010,Sasaki2011,Sasaki12,TSato13,Hosur2014,Kobayashi15PRL,Hashimoto_2015,Hashimoto_2016,Oudah2016,Kawakami2018,Kawakami2019}.
The majority of candidate superconductors have odd-parity paring and belong to class D or DIII in the ten Altland-Zirnbauer (AZ) symmetry classes~\cite{AZ1997}.
In particular, class DIII superconductors, i.e., time-reversal invariant superconductors without SU(2) spin rotation symmetry, can be topologically nontrivial in one, two, and three spatial dimensions~\cite{Schnyder08,Kitaev09,Schnyder09,Ryu10}, and their topological properties are related via the dimensional reduction~\cite{Qi2009,Qi2010}. Thus, class DIII superconductors have rich topological structures, and the materials search is pursued with the aid of the so-called Fermi surface criterion~\cite{Sato2010,Fu2010,Qi2010}.
On the other hand, superconductors of even-parity spin-singlet pairing
in class CI of the AZ symmetry classification
can have a stable topological invariant only in three spatial dimensions~\cite{Schnyder08,SchnyderPRL09}
and have received less attention than those in class DIII.

The K-theoretical classification based on stable equivalence fails to capture some topological phases. The topological phases that fall outside the K-theoretical classification are called unstable topological phases and can become topologically nontrivial only if certain conditions are met regarding the number of occupied (or total) bands or band representations. 
The search for unstable topological phases is vigorously underway and has revealed a variety of unstable topological phases such as fragile~\cite{Po2018fragile,Bouhon2020} and delicate~\cite{Nelson2021,Nelson2022} topological insulators.
For example, the Hopf insulators~\cite{Schnyder08,Moore2008,Deng2013,Kennedy2016,Liu2017,Schuster2019,Unal2019,Schuster2021,Lapierre2021,Zhu2023,Unal2020,Lim2023,jankowski2024non} are in a delicate topological phase and realized only when the total number of bands is limited to two, while the Euler insulators~\cite{Ahn2019failure,Po2019,Song2019all,Bouhon2019,Jiang2024,lee2024euler} are fragile topological insulators that are realized only when the number of valence bands is limited to two. 

Another subtle class of topological insulators that are outside of standard classification are so-called representation-protected topological phases~\cite{Alexandradinata2014spin,Alexandradinata2020,Song2020fragile,Brouwer2023},
example of which are topological insulators of class AI [with time-reversal and SU(2) spin rotation symmetries]~\cite{Fu2011,Kobayashi2021,Brouwer2023} that can be realized when the band representation is fixed to two-dimensional (2D) representations formed by non-$s$ orbital states. 
Interestingly, the stability of the representation-protected topological phases does not depend on the number of bands 
~\cite{Fu2011,Alexandradinata2020,Kobayashi2021,Kim2022three}, and in this sense they can be considered as stable topological phases \cite{Brouwer2023}.

The study of unstable topological phases has largely focused on insulators, and little is known about unstable topological phases of superconductors. A rare known example is the Hopf superconductor~\cite{Kennedy2016} in a delicate topological phase, which, however, is difficult to realize in solids due to the tight constraint on the number of bands. Lack of realistic examples hinders understanding of unstable topological superconductors. 

\begin{figure}[btp]
  \begin{center}
    \includegraphics[width=\linewidth]{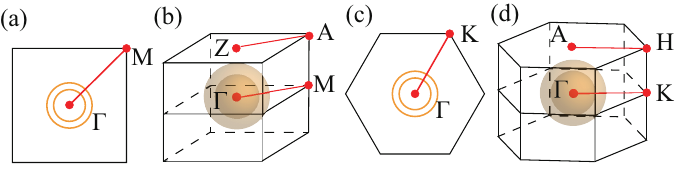}
    \caption{
    (Color online) 2D (3D) Brillouin zones of $C_4$ symmetric (a) [(b)] and $C_6$ symmetric systems (c) [(d)], where the red points indicate $C_nT$ ($n=4,6$) invariant momenta. The $\mathbb{Z}_2$ topological invariants [Eq.~(\ref{eq:c4_z2_inv}) and (\ref{eq:c6_z2_inv})] are defined on the red lines. The Fermi surface criterion insists that the $\mathbb{Z}_2$ topological invariants are nonzero if the Fermi circles (surfaces) in the 2D (3D) Brillouin zones constructed from the orbital doubles, e.g., ($d_{xz},d_{yz}$), encloses an odd number of $C_nT$ invariant momenta, and the parity of the pair potentials are opposite on them.}
    \label{fig:FS}
  \end{center}
\end{figure}    

In this paper, we propose a theoretical model of two-dimensional (2D) topological superconductors of class CI that are categorized as a representation-protected topological phase.
Specifically, we consider superconductors satisfying the following two prerequisites. (i) They have SU(2) spin rotation symmetry, time-reversal ($T$) symmetry, and $n$-fold rotation ($C_n$) symmetry with $n=4,6$. That is, the superconductors of our interest belong to class CI with additional $C_n$ symmetry.
(ii) The Fermi surfaces relevant to superconductivity are formed from a pair of non-$s$ orbital states with orbital angular momentum $2 \ell \neq 0 \mod n$, which we call an orbital doublet. For instance, orbital doublets with $\ell =1$ correspond to $(p_x,p_y)$ or $(d_{xz},d_{yz})$ orbitals and those with $\ell=2$ to $(d_{x^2 -y^2},d_{xy})$ orbitals. The condition (ii) is crucial for defining the representation-protected topological indices, since the orbital doublets behave like $S=\frac12$ spin and exhibit two-fold degeneracy at $C_nT$-invariant points in the Brillouin zone. Under the conditions (i) and (ii), a $\mathbb{Z}_2$ topological index for 2D spin-singlet superconductors is defined in terms of orbital doublets. 
In the limit of weak pair potential, the $\mathbb{Z}_2$ index is directly related to the topology of the Fermi surfaces as dictated by the following Fermi surface criterion: {\it the $\mathbb{Z}_2$ index is nonzero if the two Fermi surfaces of the orbital doublets, each enclosing one of the $C_nT$-invariant momenta, 
have $s$-wave pair potentials with opposite signs, i.e., $s_{\pm}$-wave pairing.}

Furthermore, extending the definition of the $\mathbb{Z}_2$ index to three-dimensional (3D) superconductors, we obtain a relationship between the Fermi surface topology of a 3D $s_{\pm}$-wave superconductor and a 3D winding number ($ w_{3d} \in 2\mathbb{Z}$; see Eq.~(\ref{eq:w3d})), a stable topological invariant in class CI~\cite{Schnyder08}.
This gives the following Fermi surface criterion of the 3D topological phase: {\it $w_{3d} \neq 0$ if the two Fermi surfaces of the orbital doublets, each enclosing one of the $C_nT$ invariant momenta,
have $s_{\pm}$-wave pair potentials.}
Since the 3D winding number is defined independently of the $C_n$ symmetry and band representations, the topological phases with $w_{3d}\ne0$ are robust against perturbations which break the prerequisites (i) and (ii) as long as the bulk energy gap and the defining symmetries of class CI are maintained.
We construct a model of a 3D representation-protected $s_{\pm}$-wave topological superconductor that satisfies the Fermi surface criterion, and demonstrate the existence and robustness of gapless surface states with quadratic dispersion.

Finally, we apply our theory to iron-based superconductors (FeSCs) and propose strongly hole-doped FeSC, e.g., KFe$_2$As$_2$ under pressure, to be a candidate of a topological superconductor of class CI. 

\paragraph*{Spin-singlet superconductors.} We start from a Bogoliubov-de Gennes (BdG) Hamiltonian with a spin-singlet pair potential for spin-orbit free superconductors.
With the SU(2) spin rotation symmetry, the BdG Hamiltonian is reduced to the form
\begin{align}
   H_{\sigma \sigma'}(\bm{k}) = \begin{pmatrix} \mathcal{E}_{\sigma \sigma'}(\bm{k}) & \Delta_{\sigma \sigma'}(\bm{k}) \\ \Delta^{\dagger}_{\sigma \sigma'}(\bm{k}) & -\mathcal{E}^{\ast}_{\sigma \sigma'}(-\bm{k})\end{pmatrix}\label{eq:bdg_Hami}
\end{align}
in the basis of $\Psi_{\bm{k},\sigma} =(c_{\bm{k},\uparrow,\sigma},c_{-\bm{k},\downarrow,\sigma}^{\dagger})^t$, where $c_{\bm{k},s,\sigma}$ is the annihilation operator of electron with momentum $\bm{k}$, spin $s$, and orbital $\sigma$, $\mathcal{E}(\bm{k})$ is the normal-state Hamiltonian, and $\Delta(\bm{k})$ is the spin-singlet pair potential with $\Delta(\bm{k})=\Delta^t(-\bm{k})$. The BdG Hamiltonian in Eq.~(\ref{eq:bdg_Hami}) has the particle-hole (PH) symmetry, $C H(\bm{k})C^{-1} = -H(-\bm{k})$ with $C=i\tau_y K$, where $\tau_\alpha$ ($\alpha=x,y,z$) are the Pauli matrices in the Nambu space, and $K$ is the complex conjugation. The PH operator satisfies $C^2=-1$.
We assume that the superconducting state has time-reversal and $n$-fold rotation symmetries in addition to the PH symmetry.
This means that $\mathcal{E}(\bm{k})$ and $\Delta(\bm{k})$ satisfy the symmetry relations
\begin{align}
  &U_T \mathcal{E}^{\ast}(\bm{k}) U_T^{\dagger} = \mathcal{E}(-\bm{k}),  \ \ U_T \Delta^{\ast}(\bm{k}) U_T^{t} = \Delta(-\bm{k}), \label{eq:trs}\\
  &C_n \mathcal{E}(\bm{k}) C_n^{\dagger} = \mathcal{E}(R_n\bm{k}),  \ \ C_n \Delta(\bm{k}) C_n^{t} = \Delta(R_n\bm{k}), \label{eq:cns}
\end{align}
where $U_T$ and $C_n$ are unitary matrices satisfying $U_T U_T^{\ast} =1$, $(C_n)^n=1$, and $U_TC_n^{\ast} = C_nU_T$. The $3\times3$ matrix $R_n$ represents $n$-fold rotation in the $k$ space, e.g., $R_4(k_x,k_y,k_z) = (-k_y,k_x,k_z)$.
The second equation in (\ref{eq:cns}) implies that the pair potential is invariant under the $C_n$ operation, i.e., an (extended) s-wave pairing. It is convenient to choose $U_T$ as the identity matrix, so that $\mathcal{E}(\bm{k})=\mathcal{E}^{\ast}(-\bm{k})$, $\Delta(\bm{k})=\Delta^{\dagger}(\bm{k})$, and $C_n=C_n^*$. Then the $n$-fold rotation operator can be written as $C_{n,\ell} = \exp\left(i \frac{ 2 \pi \ell}{n} \sigma_y\right)$, where $\sigma_\alpha$ ($\alpha=x,y,z$) are the Pauli matrices acting on the orbital doublet and $\ell$ is its orbital angular momentum. Hereafter, we adopt this representation unless otherwise specified.
 
Since $T^2 =1$ and $C^2=-1$ (where $T \equiv U_T K$ is the TR operator), the BdG Hamiltonian (\ref{eq:bdg_Hami}) belongs to class CI, which has stable topological phases only in three spatial dimensions, characterized by a 3D winding number ($ w_{3d} \in 2\mathbb{Z}$). In the following, we will present 2D representation-protected and 3D stable topological superconductors of class CI when the conditions (i) and (ii) are fulfilled.  

\paragraph*{2D systems.} We consider the BdG Hamiltonian of a 2D $C_4$ symmetric model of the orbital doublets ($p_x,p_y$) with $\ell=1$, for which we can explicitly define a $\mathbb{Z}_2$ topological index. 
The four-fold rotation operator for the doublet is given by $C_{4,1}=i \sigma_y$, satisfying $(C_{4,1}T)^2=-1$. The Kramers' theorem for the $C_{4,1}T$ operation indicates that the orbital doublets must be degenerate at the $C_4T$ invariant momenta $\bm{k}_{\rm inv}$, $\Gamma=(0,0)$ and $\mathrm{M}=(\pi,\pi)$, in the 2D Brillouin zone; see Fig.~\ref{fig:FS} (a). 
We choose the basis in which the chiral operator, the product of the PH and time-reversal operators, has a diagonal form, and rewrite the BdG Hamiltonian as
\begin{align}
  H(\bm{k}_\parallel) = \begin{pmatrix} 0 & A(\bm{k}_\parallel) \\ A^{\dagger}(\bm{k}_\parallel) & 0 \end{pmatrix} \label{eq:off-matrix}
\end{align}
in the basis of $(c_{\bm{k},\uparrow,\sigma}-ic^{\dagger}_{-\bm{k},\downarrow,\sigma},c_{\bm{k},\uparrow,\sigma}+ic^{\dagger}_{-\bm{k},\downarrow,\sigma})^t/\sqrt{2} $, where $\bm{k}_\parallel = (k_x,k_y)$ and $A(\bm{k}_\parallel) \equiv \mathcal{E}(\bm{k}_\parallel)-i\Delta (\bm{k}_\parallel)$.
The $C_{4,1}T$ operator acting on the off-diagonal component $A(\bm{k}_\parallel)$ satisfies the relation 
\begin{align}
  C_{4,1} A^t(\bm{k}_\parallel) = A(-R_4\bm{k}_\parallel) C_{4,1},
\end{align}
which reads $ A(\bm{k}_{\rm inv})C_{4,1}=- [ A(\bm{k}_{\rm inv}) C_{4,1}]^t$ at $\bm{k}=\bm{k}_\mathrm{inv}$ because $C_{4,1}^t=-C_{4,1}$.
The skew-symmetric property of $A(\bm{k}_{\rm inv})C_{4,1}$ allows us to define a $\mathbb{Z}_2$ topological invariant $\nu_4$ from $A(\bm{k}_\parallel)$ on the $\Gamma$-M line, 
\begin{align}
   (-1)^{\nu_4} = \frac{\pf[A(\mathrm{M})C_{4,1}]}{\pf[A(\Gamma)C_{4,1}]} 
   \exp \!\left(\int^{\Gamma}_\mathrm{M} \frac{dk}{2} \, \partial_k \log \{\det[A(\bm{k})]\}\right)\! . \label{eq:c4_z2_inv}
\end{align}
The expression~(\ref{eq:c4_z2_inv}) is gauge invariant and only takes the values $\pm 1$~\cite{Qi2010,suppl}. Thus, a spin-singlet superconducting state with $\nu_4=1$ cannot be continuously transformed to any spin-singlet superconductor with $\nu_4=0$ without closing the superconducting gap. Since Eq.~(\ref{eq:c4_z2_inv}) is defined solely in terms of the orbital doublets, hybridization with an $s$-orbital band makes Eq.~(\ref{eq:c4_z2_inv}) ill-defined. The topological superconductor of orbital doublets is a superconducting analog of the class AI representation-protected topological insulator proposed by Fu~\cite{Fu2011}.
 
Equation (\ref{eq:c4_z2_inv}) can be simplified in the weak-coupling limit~\cite{Qi2010} where $|\Delta|$ is much smaller than the Fermi energy. Assuming that $\bm{k}_{\rm inv}$s are not on the Fermi surface, we evaluate the change in $\theta_{\bm{k}_\parallel} \equiv \arg[\det(A(\bm{k}_\parallel))]$ along the $\Gamma$-M line. Since $\Delta(\bm{k}_\parallel)$ is nonzero only in the vicinity of the Fermi surfaces, we can approximate $A(\bm{k}_\parallel)$ as $(\epsilon_{n,\bm{k}_\parallel} -i \delta_{n,\bm{k}_\parallel}) | u_{n,\bm{k}_\parallel}\rangle \langle u_{n,\bm{k}_\parallel}|$, where $\epsilon_{n,\bm{k}_\parallel}$ and $| u_{n,\bm{k}_\parallel}\rangle$ are the $n$th eigenvalue and eigenstate of $\mathcal{E}(\bm{k}_\parallel)$, and
$\delta_{n,\bm{k}_\parallel} = \langle u_{n,\bm{k}_\parallel} |\Delta(\bm{k}_\parallel)| u_{n,\bm{k}_\parallel} \rangle \in\mathbb{R}$. For $\bm{k}_\parallel$ away from the Fermi surfaces, $\theta_{\bm{k}_\parallel}$ can take only $0$ or $\pm \pi$ since $\delta_{n,\bm{k}_\parallel}=0$. Thus, $\theta_{\bm{k}_\parallel}$ changes suddenly at the Fermi surfaces, and its variation $\pm \pi$ across a Fermi surface is determined from the sign of $\delta_{n,\bm{k}_\parallel}$ on the Fermi surface. Collecting the phase jumps at all Fermi surfaces, labeled by $s=1,\dots, m $, crossing the $\Gamma$-M line, we evaluate the $\mathbb{Z}_2$ invariant as
\begin{align}
(-1)^{\nu_4} = \prod_{s=1}^m \sgn(\delta_{s}),  \label{eq:nu4-FS}
\end{align}
where $\delta_s \equiv \delta_{s,\bm{k}_\parallel}$ is the pair potential on the $s$th Fermi surface. Here, the Pfaffian factors in Eq.\ (\ref{eq:c4_z2_inv}) are canceled out by the contributions other than the phase jumps discussed above from the integral. See Supplementary materials (SM)~\cite{suppl} for more details. Equation~(\ref{eq:nu4-FS}) gives $\nu_4 = 1$ when the pair potential has different signs on the two Fermi surfaces as shown for the $s_{\pm}$-wave superconductor in Fig.~\ref{fig:FS}(a).
We note that the number of the Fermi surfaces crossing the $\Gamma$-M line is restricted to be an even integer by the two-fold degeneracy at the $\Gamma$ and M points.
Unfortunately, the topological protection of gapless edge states is absent in two dimensions since $C_4$ symmetry is not preserved at the boundaries. A model of a 2D $s_{\pm}$-wave topological superconductor is discussed in SM~\cite{suppl}, where the edge states are shown to have a tiny energy gap.

\begin{figure}[tbp]
  \begin{center}
    \includegraphics[width=\linewidth]{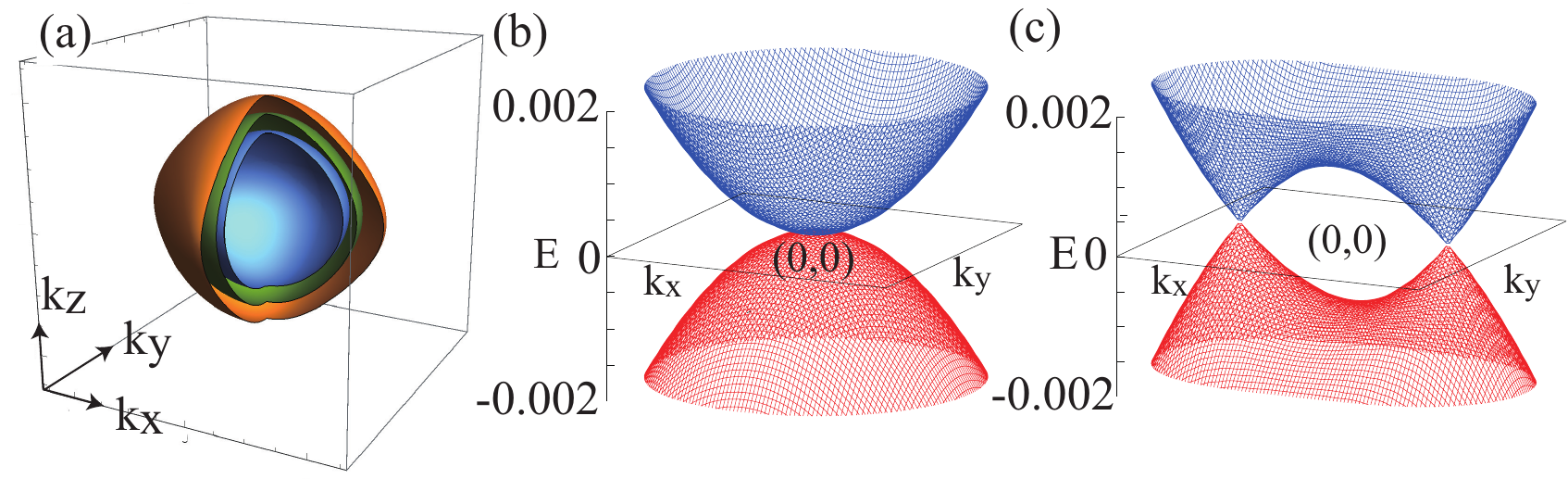}
    \caption{
    (Color online) (a) The inner (blue), middle (green), and outer (yellow) spherical surfaces, respectively, represent the inner Fermi surface, the zeros of the pair potential, and the outer Fermi surface of the tight-binding model (\ref{eq:3d_model}), where we cut a quadrant to visualize the inner surfaces. The parameters are chosen as $(t,v_1,v_2,v_3,\mu)=(1,0.3,0.3,0.3,2)$ and $(\Delta_0,\Delta_1) =(0.2,-0.1)$. (b) The energy dispersion of the (001) surface states in the $k_x$-$k_y$ plane. (c) The dispersion of the (001) surface states when a $C_4$ breaking perturbation $0.01 \sigma_x$ is added to the normal-state Hamiltonian.  }
    \label{fig:surface}
  \end{center}
\end{figure}

\paragraph*{3D systems.}
Let us assume that 3D superconductors of our interest have $C_4$ rotation symmetry along the $z$ axis.
The 3D systems have two $C_4T$ symmetric planes $k_z=0$ and $\pi$, on which we can define the $\mathbb{Z}_2$ topological invariants independently, labeled by $\nu_4(0)$ and $\nu_4(\pi)$. The $C_4T$ invariant momenta are shown in Fig.~\ref{fig:FS} (b). The $\nu_4(0)$ and $\nu_4(\pi)$ are defined on the $\Gamma$-M line and the Z-A line, respectively. Then, 3D topological phases are characterized by the difference $\tilde{\nu}_4 = \nu_4(\pi) - \nu_4(0) \mod 2$. We call the phase with $\tilde{\nu}_4 =1$ a strong topological phase. The strong topological phase has a gapless topological surface mode on the (001) surface that is invariant under the $C_4$ symmetry. On the other hand, when $\tilde{\nu}_4=0$, we have either a trivial phase or a weak topological phase, i.e., a stack of 2D topological phases.

Furthermore, there are stable topological phases characterized by the 3D winding number $w_{3d}$ in class CI. The 3D winding number is defined as follows~\cite{Schnyder08}. The BdG Hamiltonian of a 3D class CI superconductor can be transformed to the off-diagonal matrix form as in Eq.~(\ref{eq:off-matrix}). Then, $A(\bm{k})$ is rewritten by singular value decomposition as $A(\bm{k}) = U(\bm{k}) D(\bm{k}) V^{\dagger}(\bm{k})$ with the unitary matrices $U(\bm{k})$ and $V(\bm{k})$, where $\bm{k} =(k_x,k_y,k_z)$. Since $D(\bm{k})$ is positive definite, we can deform it to the identity matrix without closing the energy gap. Therefore, we can replace $A(\bm{k})$ with a unitary matrix $Q(\bm{k}) = U(\bm{k})V^{\dagger}(\bm{k})$ and define the 3D winding number 
\begin{align}
    w_{\rm 3d} = &\int \frac{d^3k}{24 \pi^2} \sum_{\alpha, \beta, \gamma} \epsilon^{\alpha \beta \gamma} \tr\Big\{
    \left[Q^{\dagger}(\bm{k}) \partial_{k_\alpha} Q(\bm{k})\right] \nonumber \\
    &\times \left[Q^{\dagger}(\bm{k}) \partial_{k_\beta} Q(\bm{k})\right] \!
    \left[Q^{\dagger}(\bm{k}) \partial_{k_\gamma} Q(\bm{k})\right] \Big\}, \label{eq:w3d}
\end{align}
where $\alpha, \beta, \gamma=x,y,z$ and $\epsilon^{\alpha \beta \gamma}$ is the Levi-Civita symbol. The winding number $w_{\rm 3d}$ takes even-integer values in class CI.
In the weak-coupling limit, it is reduced to the form $w_{\rm 3d} = \frac{1}{2} \sum_s \sgn (\delta_{k_F,s}) C_s $~\cite{Qi2010}, where $C_s$ is the Chern number of the $s$th Fermi surface, and the summation is taken over all the Fermi surfaces. Using this Fermi surface formula,
we find that $\tilde{\nu}_4$ and $w_{\rm 3d}$ are related by
 \begin{align}
  w_{\rm 3d}=2\tilde{\nu}_4 \mod 4. \label{eq:fs_c4}
 \end{align}
 The derivation of Eq.~(\ref{eq:fs_c4}) is given in SM~\cite{suppl}. Equation (\ref{eq:fs_c4}) is one of our main results, bridging the representation-protected $\mathbb{Z}_2$ topology and the stable topology. 
Note that the formula is meaningful
only in the absence of inversion symmetry, 
since the inversion symmetry leads to $w_{\rm 3d} =0$~\cite{Qin2022}. 

A tight-binding model of a 3D $s_{\pm}$-wave superconductor from ($p_x,p_y$) or ($d_{xz},d_{yz}$) orbital pairs is given by
\begin{subequations}
\label{eq:3d_model}
\begin{align}
\mathcal{E}(\bm{k}) = \,& t[\cos(k_x) +\cos(k_y) + \cos(k_z)]- \mu \notag \\
&+ v_1 \sin(k_x) \sin(k_y) \sigma_z
 \notag \\
& + v_2[\cos(k_x)
  -\cos(k_y)]\sigma_x + v_3 \sin(k_z) \sigma_y, \\
\Delta(\bm{k}) =\, & \Delta_0 + \Delta_1 [\cos(k_x)+\cos(k_y)+\cos(k_z)].
\end{align}
\end{subequations}
Equation~(\ref{eq:3d_model}) is invariant under the time-reversal ($T=K$) and $C_4$ transformations ($C_4=i \sigma_y$). Here, the $v_3$ term breaks inversion symmetry (the inversion operator is the identity matrix). 
The Fermi surfaces and the zeros of pair potential are shown in Fig.~\ref{fig:surface} (a), where we have chosen the parameters such that the Fermi surfaces enclose the $\Gamma$ point and the pair potential changes its sign between the two Fermi surfaces, whereby leading to $\tilde{\nu}_4 = 1$. Figure~\ref{fig:surface} (b) shows gapless surface states with quadratic dispersion on the (001) surface. Here, the quadratic dispersion is attributed to the Kramers degeneracy of the orbital doublet enforced by the $C_4T$ symmetry. When a $C_4$ breaking perturbation is introduced, the quadratic surface mode is split, as shown in Fig.~\ref{fig:surface} (c), into two gapless Dirac cones with linear dispersion that are protected by the stable 3D winding number.

\paragraph*{Generalization to $C_6$ symmetric cases.} We now turn to $C_6$ symmetric cases. The 2D and 3D Brillouin zones are shown in Fig.~\ref{fig:FS} (c) and (d), where $C_6T$ invariant momenta are denoted by $\Gamma$, Z, K, and H. After repeating the same procedure as the $C_4$ symmetric case, we obtain the weak $\mathbb{Z}_2$ topological invariant as
 \begin{align}
   (-1)^{\nu_{6,\ell}(0)} =& \frac{\pf[A(\mathrm{K})C^a_{6,\ell} ]}{\pf[A(\Gamma)C^a_{6,\ell} ]}
   \exp \!\left( \int^{\Gamma}_\mathrm{K} \!\frac{dk}{2} \, \partial_k \log \{\det[A(\bm{k})]\}\right)
   \label{eq:c6_z2_inv}
 \end{align}
for orbital doublets $\ell=1$ and 2 of 2D representations of the six-fold rotation operators.
We have defined $C^a_{6,\ell} \equiv C_{6,\ell} - C_{6,\ell}^t$ to extract the skew-symmetric part of $C_{6,\ell}$.
The $\mathbb{Z}_2$ index for the $k_z=\pi$ plane $\nu_{6,\ell}(\pi)$ is obtained by replacing $\Gamma$ and K with Z and H. Using Eq.~(\ref{eq:c6_z2_inv}), the strong $\mathbb{Z}_2$ topological index is defined by $\tilde{\nu}_{6,\ell} \equiv \nu_{6,\ell}(\pi) - \nu_{6,\ell}(0)$. 

In the weak-coupling limit, Eq.~(\ref{eq:c6_z2_inv}) reduces to Eq.~(\ref{eq:nu4-FS}), where we take the product of $\sgn(\delta_s)$ along the $\Gamma$-K line for $\nu_{6,\ell}(0)$ and along the Z-H line for $\nu_{6,\ell}(\pi)$. Thus, $\nu_{6,\ell}$ can also take a nonzero value for an $s_{\pm}$-wave superconductor.

\paragraph*{Discussion.}
Finally, we discuss an application to $s_{\pm}$-wave paring states in FeSCs. After the discovery of the superconductivity of F-doped LaFeAsO in 2008~\cite{Kamihara2008}, several related families of FeSCs have been discovered~\cite{Ishida2009,Paglione2010,Johnston2010,Hirschfeld2011,Stewart2011,Kordyuk2012,Chen2014,Hosono2015}. The superconducting states of FeSCs are expected from their high $T_{\rm c}$ to be an unconventional pairing such as $d$-wave and $s_{\pm}$-wave pairings, and the antiferromagnetic spin fluctuation theory supports an $s_{\pm}$ pairing~\cite{Kuroki2008,Mazin2008}. The basic crystal structure of the FeSCs has the P4/nmm symmetry, i.e., $C_4$ symmetry is preserved. Due to their layer structure, the Fermi surfaces of FeSCs have quasi-2D shape. They comprise small Fermi pockets surrounding high-symmetry momenta, and, in particular, there are commonly two hole pockets around the $\Gamma$ point and two electron pockets around the $M$ point. These Fermi pockets are formed from the $d$ orbitals of the Fe atoms: the hole pockets from $d_{xz}$/$d_{yz}$ orbitals, and the electron pockets from $d_{xz}$, $d_{yz}$, and $d_{xy}$ orbitals~\cite{Hirschfeld2011,Hosono2015}. Hence, for compounds of the FeSCs family with only hole pockets, the conditions (i) and (ii) are satisfied if the effect of spin-orbit interaction on superconducting states is weak. When $s_{\pm}$-wave pairing is formed within the two hole pockets, the $\mathbb{Z}_2$ index becomes nontrivial as shown in Fig.~\ref{fig:FS} (a), and thus a 2D representation-protected topological phase is realized.

An example of the materials with only hole pockets is KFe$_2$As$_2$~\cite{Rotter2008,Sasmal2008}, the end member of hole-doped (Ba$_{1-x}$K$_x$)Fe$_2$As$_2$ series, where the hole doping wipes out the electron pockets~\cite{Sato2009}. The main hole pockets contributing to superconductivity are formed from the $d_{xz}$/$d_{yz}$ orbitals~\footnote{An additional $d_{xy}$ orbital hole pocket also arises around the $\Gamma$ point with a negligibly small superconducting gap~\cite{Okazaki2012octet}. }. The laser ARPES measurements~\cite{Okazaki2012octet,wu2024nodal} have reported that superconducting gap nodes exist on one of the hole pockets, which suggests that the superconductivity in KFe$_2$As$_2$ is nodal $s_{\pm}$ wave. Thus, this material can be considered to be located at the topological phase boundary. In addition, resistivity measurements on KFe$_2$As$_2$ under pressure have found that the pressure dependence of $T_{\rm c}$ changes from negative to positive at the critical pressure $P_{\rm c} \sim 18 $ kbar without changing the Fermi surface topology~\cite{Tafti2013,Terashima2014}, indicating that a crossover from the nodal $s_{\pm}$ wave to a full-gap $s$ wave occurs. 
Therefore, pressurized KFe$_2$As$_2$ is expected to be a good test bed to study the representation-protected topology.

We expect from the Fermi surface criterion that materials with only electron pockets, such as KFe$_2$Se$_2$~\cite{Qian2011absence,Zhang2011nodeless,Dagotto2013}, monolayer FeSe~\cite{Wang2012interface,Liu2012electronic,Wang2017high,Coldea2018,Kreisel2020}, and LiFe(OH)FeSe~\cite{Niu2015,Zhao2016common}, should also be potential candidates. Investigating representation-protected topological phases in other FeSCs is also an interesting future work.

\paragraph*{Summary.}
We have shown that the representation-protected $\mathbb{Z}_2$ topological invariants [Eqs.~(\ref{eq:c4_z2_inv}) and (\ref{eq:c6_z2_inv})] are defined for fully-gapped superconductors with SU(2) spin rotation, $C_n$ rotation ($n=4,6$), and time-reversal symmetries, when the Fermi surfaces consist only of the orbital doublets with the angular momentum $\ell=1$ for $C_4$ and $\ell=1,2$ for $C_6$. We have  established the relation between the $\mathbb{Z}_2$ index and the 3D winding number via the Fermi-surface formula (\ref{eq:nu4-FS}) and (\ref{eq:fs_c4}) in the weak-coupling limit. We have presented a model of 3D topological $s_{\pm}$ superconductors, which has a gapless surface mode with quadratic dispersion. Such gapless surface mode would be observed through surface-sensitive measurements, such as tunneling spectroscopy~\cite{Hsieh2012,Yamakage_2012}, the Josephson effect~\cite{Fu-Kane2009,Yamakage2013}, and penetration depth measurement~\cite{Wu2020power}. Finally, we have proposed pressurized KFe$_2$As$_2$ as a potential candidate of a representation-protected topological superconductor in class CI.

S.K. is very grateful to K. Shiozaki, the discussion with whom led to the initial idea of the paper. S.K. also thanks M. Hirayama for useful discussions. This work was supported by JSPS KAKENHI (Grants No.\ JP19K03680, No.\ JP19K14612, No.\ JP19H01824, and No.\ JP22K03478) and JST CREST (Grant No.\ JPMJCR19T2).

\bibliography{spin_singlet_tsc}

\end{document}


\preprint{APS/123-QED}

\renewcommand{\thefigure}{S\arabic{figure}} 

\renewcommand{\thetable}{S\arabic{table}} 

\renewcommand{\thesection}{S\arabic{section}.}

\renewcommand{\theequation}{S.\arabic{equation}}

\setcounter{figure}{0}
\setcounter{table}{0}
\setcounter{equation}{0}


\title{Supplementary materials: \\ Representation-protected topology of spin-singlet $s$-wave superconductors}
\author {Shingo Kobayashi}
\affiliation{RIKEN Center for Emergent Matter Science, Wako, Saitama, 351-0198, Japan}
\author{Akira Furusaki}
\affiliation{RIKEN Center for Emergent Matter Science, Wako, Saitama, 351-0198, Japan}
\affiliation{Condensed Matter Theory Laboratory, RIKEN, Wako, Saitama, 351-0198, Japan}
\date{\today}

\maketitle


\begin{flushleft} 
{\bf S1. Bogoliubov de-Gennes Hamiltonian of spin-singlet superconductors }
\end{flushleft}

In this section, we show that in the presence of the SU(2) spin rotation (SR) symmetry, a superconducting state is described by the Bogoliubov de-Gennes (BdG) Hamiltonian (1) in the main text. We start from the general form of the BdG Hamiltonian, 
$H_{\rm BdG}=\frac{1}{2} \sum_{\bm{k},s,s',\sigma,\sigma'} \Psi_{\bm{k},s,\sigma}^\dagger \widetilde{H}_{s,\sigma;s'\sigma'}(\bm{k}) \Psi_{\bm{k},s',\sigma'}$ with
$\Psi_{\bm{k},s,\sigma} =(c_{\bm{k},s,\sigma},c_{-\bm{k},s,\sigma}^{\dagger})^t$ and
 \begin{align}
   \widetilde{H}_{s,\sigma; s', \sigma'}(\bm{k}) = \begin{pmatrix} \tilde{\mathcal{E}}_{s,\sigma ;s',\sigma'}(\bm{k}) & \tilde{\Delta}_{s,\sigma ;s',\sigma'}(\bm{k}) \\ \tilde{\Delta}^{\dagger}_{s,\sigma ;s',\sigma'}(\bm{k}) & -\tilde{\mathcal{E}}^{\ast}_{s,\sigma ;s',\sigma'}(-\bm{k})\end{pmatrix}_\tau, \label{eq:bdg_Hami}
 \end{align}
 where $c_{\bm{k},s,\sigma}$ is the annihilation operator of electron with momentum $\bm{k}$, spin $s$, orbital $\sigma$, the normal-state Hamiltonian $\tilde{\mathcal{E}}(\bm{k})$ is a Hermitian matrix , and the gap function $\tilde{\Delta}(\bm{k})$ satisfies $\tilde{\Delta}(\bm{k})=-\tilde{\Delta}^t(-\bm{k})$ due to the Fermi-Dirac statistics. The subscript $\tau$ of the matrix implies that the matrix is represented in the Nambu basis. We suppose that the BdG Hamiltonian satisfies the particle-hole (PH), time-reversal (TR), and SR symmetries as follows.
 \begin{align}
     \text{PH symmetry}:& \ \ C \widetilde{H}(\bm{k}) C^{-1} = -\widetilde{H}(-\bm{k}), \ \ C=\tau_x K, \label{eq:phs} \\
     \text{TR symmetry}:& \ \ T \widetilde{H}(\bm{k}) T^{-1} = \widetilde{H}(-\bm{k}), \ \ T=i s_y K, \label{eq:trs} \\
     \text{SR symmetry}:& \ \ \mathcal{S} \widetilde{H}(\bm{k}) \mathcal{S}^{-1} = \widetilde{H}(\bm{k}), \ \ \mathcal{S} = \begin{pmatrix} e^{i \bm{s} \cdot \bm{\theta}/2} & 0 \\ 0 & e^{-i \bm{s}^{\ast} \cdot \bm{\theta}/2} \end{pmatrix}_\tau, \label{eq:srs}
 \end{align}
where $s_i$ and $\tau_i$ are the Pauli matrices in the spin and Numbu spaces and $K$ is the complex conjugation. $\bm{\theta}=(\theta_x,\theta_y,\theta_z)$ represent the rotation angle around the $x$, $y$, $z$ axes in the spin space. Taking the derivative of Eq.~(\ref{eq:srs}) with respect to $\bm{\theta}$ yields the commutation relations with the generator $\mathcal{J}_i$ ($i=x,y,z$)~\cite{SatoFujimoto16},
\begin{align}
    [\mathcal{J}_i, \widetilde{H}(\bm{k})]=0, \ \ \mathcal{J}_i = \begin{pmatrix} s_i & 0 \\ 0 & -s_i^{\ast} \end{pmatrix}_\tau.
\end{align}
Thus, the BdG Hamiltonian is block-diagonalized in the basis diagonalizing one of $\mathcal{J}_i$'s, say $\mathcal{J}_z$. Each block matrix has the eigenvalue $\mathcal{J}_z = \pm 1$. In addition, using the other generators $\mathcal{J}_x$ and $\mathcal{J}_y$, we find that the block matrix with $\mathcal{J}_z =  1$ is related to that with $\mathcal{J}_z =  -1$ since $\{ \mathcal{J}_i, \mathcal{J}_j \} =0$ ($i \neq j$). Thus, the BdG Hamiltonian (\ref{eq:bdg_Hami}) is reduced to the block matrix with $\mathcal{J}_z = 1$, whose basis is given by $\Psi_{\bm{k},\sigma} =(c_{\bm{k},\uparrow,\sigma},c_{-\bm{k},\downarrow,\sigma}^{\dagger})^t$. The reduced BdG Hamiltonian is given by 
\begin{align}
   H_{\sigma \sigma'}(\bm{k}) = \begin{pmatrix} \mathcal{E}_{\sigma \sigma'}(\bm{k}) & \Delta_{\sigma \sigma'}(\bm{k}) \\ \Delta^{\dagger}_{\sigma \sigma'}(\bm{k}) & -\mathcal{E}^{\ast}_{\sigma \sigma'}(-\bm{k})\end{pmatrix}_\tau, \label{eq:bdg_mini}
 \end{align}
 where we choose the representations of $\tilde{\mathcal{E}}_{\sigma ;\sigma'}$ and $\tilde{\Delta}_{\sigma ;\sigma'}$ as
\begin{align}
    &\tilde{\mathcal{E}}_{\sigma ;\sigma'} = \begin{pmatrix} \mathcal{E}_{\sigma \sigma'}(\bm{k}) & 0 \\ 0 & \mathcal{E}_{\sigma \sigma'}(\bm{k}) \end{pmatrix}_s, \label{eq:normal_state}\\
    &\tilde{\Delta}_{\sigma ;\sigma'} = \begin{pmatrix} 0 & \Delta_{\sigma \sigma'}(\bm{k}) \\ -\Delta_{\sigma \sigma'}(\bm{k}) & 0 \end{pmatrix}_s, \label{eq:gap_func} 
\end{align}
 so as to satisfy the SR symmetry (\ref{eq:srs}). Equations (\ref{eq:normal_state}) and (\ref{eq:gap_func}) represent a spin-orbit free normal-state Hamiltonian and a spin-singlet pair potential, respectively. Here, the subscript $s$ of the matrix implies the matrix representation in the spin space, and the pair potential satisfies 
\begin{align}
    \Delta(\bm{k}) = \Delta^t(-\bm{k}).
    \label{eq:Delta^t}
\end{align}

 We now consider the symmetry properties of Eq.~(\ref{eq:bdg_mini}). Since $\{\mathcal{J}_z,C\}=\{\mathcal{J}_z,T\}=0$, Eq.~(\ref{eq:bdg_mini}) is not invariant under the operation of $C$ and $T$. In fact, the operation of $C$ and $T$ exchanges the $\mathcal{J}_z =1$ state to the $\mathcal{J}_z =-1$ state. Thus, instead of using $C$ and $T$, we define modified PH and TR operators as $C' \equiv J_y C$ and $T' \equiv J_y T$. Since $C'$ and $T'$ commute with $\mathcal{J}_z$, they are preserved in Eq.~(\ref{eq:bdg_mini}). In addition, they satisfy $(C')^2=-\mathbbm{1}$, $(T')^2=\mathbbm{1}$, and $[C',T']=0$, implying that the BdG Hamiltonian belongs to class CI. Here, $\mathbbm{1}$ is the identity matrix.
 If the BdG Hamiltonian (\ref{eq:bdg_Hami}) has the $n$-fold rotation symmetry additionally, it can be preserved in the block-diagonalized Hamiltonian~(\ref{eq:bdg_mini}) since we can choose a spin-independent representation of the $n$-fold rotation under the SR symmetry.

\begin{flushleft} 
{\bf S2. Derivation of the Fermi surface formula of the $\mathbb{Z}_2$ invariant}
\end{flushleft}

In this section, we prove the Fermi surface formula of the $\mathbb{Z}_2$ invariant (7) in the main text. The proof is based on Appendix D in Ref.~\onlinecite{Qi2010}. We revisit the argument in a spin-singlet superconductor with $C_4$ rotation and TR symmetries. We focus our attention to two-dimensional (2D) systems and begin with the BdG Hamiltonian [Eq.~(1) in the main text],
\begin{align}
   H_{\sigma \sigma'}(\bm{k}) = \begin{pmatrix} \mathcal{E}_{\sigma \sigma'}(\bm{k}) & \Delta_{\sigma \sigma'}(\bm{k}) \\ \Delta^{\dagger}_{\sigma \sigma'}(\bm{k}) & -\mathcal{E}^{\ast}_{\sigma \sigma'}(-\bm{k})\end{pmatrix}_\tau, \label{eq:bdg_Hamiv2}
 \end{align}
 which satisfies the PH symmetry, TR symmetry, and $C_4$ rotation symmetry such that
\begin{align}
     & C H(k_x,k_y) C^{-1} = -H(-k_x,-k_y), \ \ C= i\tau_y K, \label{eq:phsv2} \\
     &T H(k_x,k_y) T^{-1} = H(-k_x,-k_y), \ \ T= \begin{pmatrix} U_T & 0 \\ 0 &U_T^{\ast}  \end{pmatrix}_\tau K, \label{eq:trsv2} \\
      &\tilde{C}_4 H(k_x,k_y) \tilde{C}_4^{-1} = H(-k_y,k_x), \ \ \tilde{C}_4= \begin{pmatrix} C_4 & 0 \\ 0 &C_4^{\ast}  \end{pmatrix}_\tau, \label{eq:c4s} 
\end{align}
where $K$ is the complex conjugation and $U_T$ and $C_4$ are unitary matrices satisfying $U_T U_T^{\ast} =\mathbbm{1}$ (i.e., $U_T=U_T^t$), $(C_4)^4=\mathbbm{1}$, and $U_TC_4^{\ast} = C_4U_T$. In addition, we assume $(\tilde{C}_4T)^2 = - \mathbbm{1}$, which leads to $C_4U_T = - C_4^{\dagger} U_T^t$. This assumption implies that we choose the $C_4$ representation of the orbital doublets with orbital angular momentam $\ell =1$. Equations (\ref{eq:trsv2}) and (\ref{eq:c4s}) are written as
\begin{align}
  &U_T \mathcal{E}^{\ast}(k_x,k_y) U_T^{\dagger} = \mathcal{E}(-k_x,-k_y),  \ \ U_T \Delta^{\ast}(k_x,k_y) U_T^{t} = \Delta(-k_x,-k_y) \label{eq:trsv2-2}\\
  &C_4 \mathcal{E}(k_x,k_y) C_4^{\dagger} = \mathcal{E}(-k_y,k_x),  \ \ C_4 \Delta(k_x,k_y) C_4^{t} = \Delta(-k_y,k_x). \label{eq:c4s-2}
\end{align}
In addition, the BdG Hamiltonian is invariant under the combination of PH and TR operators (chiral symmetry) defined by 
\begin{align}
     U_{\Gamma} \equiv -i CT = 
    \begin{pmatrix}
    0 & - i U_T \\
    i U_T^{\ast} & 0 
    \end{pmatrix}_\tau,
\end{align}
where we have introduced the factor $-i$ to satisfy $(U_{\Gamma})^2 = \mathbbm{1}$. We choose the basis diagonalizing $U_{\Gamma}$, which is given by applying a unitary matrix,
\begin{align}
    V = \frac{1}{\sqrt{2}}
    \begin{pmatrix}
        \mathbbm{1} & -\mathbbm{1} \\ \mathbbm{1} & \mathbbm{1}
    \end{pmatrix}_\tau
    \begin{pmatrix}
        \mathbbm{1} & 0 \\ 0 & i U_T
    \end{pmatrix}_\tau.
\end{align}
In this basis, $U_{\Gamma}$, $H(k_x,k_y)$, and $\tilde{C}_4T$ are transformed to
\begin{align}
    &VU_{\Gamma} V^{\dagger} =   
    \begin{pmatrix}
        \mathbbm{1} & 0 \\ 0 & -\mathbbm{1}
    \end{pmatrix}_\tau, \\
    &VH(k_x,k_y)V^{\dagger} =   
    \begin{pmatrix}
        0 & A(k_x,k_y)  \\ A^{\dagger}(k_x,k_y)  & 0
    \end{pmatrix}_\tau, \label{uni_Hami}\\
    &V(\tilde{C}_4T)V^{\dagger} =   
    \begin{pmatrix}
        0 &  C_4U_T \\ C_4U_T  & 0
    \end{pmatrix}_\tau  K ,\label{uni_c4t}
\end{align}
where 
\begin{align}
    A(k_x,k_y) \equiv \mathcal{E}(k_x,k_y) - i U_T \Delta^{\dagger}(k_x,k_y),
\end{align}
and we have used the relation $\Delta(k_x,k_y)U_T^{\dagger} = U_T \Delta^{\dagger}(k_x,k_y)$ that follows from Eqs.~(\ref{eq:Delta^t}) and (\ref{eq:trsv2-2}).
The $C_4T$ symmetry of the BdG Hamiltonian is also rewritten, using Eqs.~(\ref{eq:trsv2}), (\ref{eq:c4s}), (\ref{uni_Hami}), and (\ref{uni_c4t}), as
\begin{align}
    C_4 U_T A^t(k_x,k_y) = A(k_y,-k_x)C_4 U_T.
\end{align}
 At the $C_4T$ invariant momenta $\bm{k}_{\rm inv}$, i.e., $\Gamma=(0,0)$ and $M=(\pi,\pi)$ in the 2D Brillouin zone, we obtain
\begin{align}
 &C_4 U_T A^t(\bm{k}_{\rm inv}) = A(\bm{k}_{\rm inv})C_4 U_T \label{eq:c4t-A} \\
 \Longrightarrow & \ \  A(\bm{k}_{\rm inv})C_4 U_T =-[A(\bm{k}_{\rm inv})C_4 U_T]^t , \label{eq:c4t-A2} 
\end{align}
where we have used $[A(\bm{k}_{\rm inv})C_4 U_T]^t = U_T^t C_4^t A^t(\bm{k}_{\rm inv}) = C_4^\dagger U_T^t A^t(\bm{k}_{\rm inv}) = - C_4 U_T A^t(\bm{k}_{\rm inv}) $ in the second line. Therefore, $A(\bm{k}_{\rm inv})C_4 U_T$ is a skew-symmetric matrix. Using this property, the $\mathbb{Z}_2$ invariant is defined by
\begin{align}
  (-1)^{\nu_4} =& \frac{\pf[A(M)C_4 U_T]}{\pf[A(\Gamma)C_4 U_T]} \exp \left\{ \frac{-1}{2} \int_{\Gamma}^M dk \, \partial_k \log [\det(A(k_x,k_y))]\right\}, \label{eq:c4_inv}
\end{align}
where the integral is calculated along the straight path connecting the $\Gamma$ and M points. 
Furthermore, the chiral component $A(k_x,k_y)$ can be decomposed by singular value decomposition as $A(k_x,k_y)=U^{\dagger}(k_x,k_y)D(k_x,k_y)V(k_x,k_y)$, where $U(k_x,k_y)$ and $V(k_x,k_y)$ are unitary matrices and $D(k_x,k_y)$ is a diagonal matrix with non-negative elements. For a fully-gapped superconductor, we can adiabatically deform $D(k_x,k_y)$ to the identity matrix without closing the superconducting gap since it is positive definite. After the deformation, $A(k_x,k_y)$ is given by a unitary matrix $Q(k_x,k_y) = U^{\dagger}(k_x,k_y)V(k_x,k_y) $. Since the topological number remains unchanged under the deformation, Eq.~(\ref{eq:c4_inv}) is recast into
\begin{align}
    (-1)^{\nu_4} =& \frac{\pf[Q(M)C_4 U_T]}{\pf[Q(\Gamma)C_4 U_T]} \exp \left\{ \frac{-1}{2} \int_{\Gamma}^M dk \, \partial_k \log [\det(Q(k_x,k_y))]\right\}. \label{eq:c4_inv2}
\end{align}
The refined formulation is useful to check some basic properties of the $\mathbb{Z}_2$ invariant. First, we show that $\nu_4$ only takes $0$ or $1$ mod 2. Since $Q(k_x,k_y)$ is a unitary matrix, $\det[Q(k_x,k_y)] = \exp[i \phi (k_x,k_y)]$. Thus, the square of the exponential factor of Eq.~(\ref{eq:c4_inv2}) is calculated as
\begin{align}
    \exp \left\{ - \int_{\Gamma}^M dk \, \partial_k \log [\det(Q(k_x,k_y))]\right\} = e^{-i [\phi(M)-\phi(\Gamma)]} = \frac{\det[Q(\Gamma)C_4 U_T]}{\det[Q(M)C_4 U_T]}, 
\end{align}
where we have used the identity $\det[Q(\bm{k}_{\rm inv})C_4 U_T]=\det[Q(\bm{k}_{\rm inv})]\det[C_4 U_T]$. Therefore, we obtain 
\begin{align}
    (-1)^{2\nu_4} = \frac{\det[Q(M)C_4 U_T]}{\det[Q(\Gamma)C_4 U_T]} \exp \left\{ - \int_{\Gamma}^M dk \, \partial_k \log [\det(Q(k_x,k_y))]\right\} = 1,
\end{align}
which leads to $2 \nu_4 = 0 \mod 2$. Next, we show that the formula is gauge invariant. We consider an infinitesimal deformation $Q' (k_x,k_y) = Q (k_x,k_y) + \delta Q (k_x,k_y)$, which yields
\begin{align}
    \det[Q'(k_x,k_y)] = \exp[i\phi'(k_x,k_y)]=\exp\{i[\phi(k_x,k_y) + \delta \phi(k_x,k_y)]\} \label{eq:infini_Q}
\end{align}
 with $\delta \phi(k_x,k_y) \equiv \phi'(k_x,k_y) - \phi(k_x,k_y)$. Thus,
 \begin{align}
     \exp \left\{ \frac{-1}{2} \int_{\Gamma}^M dk \, \partial_k \log [\det(Q'(k_x,k_y))]\right\} =  \exp \left\{ \frac{-1}{2} \int_{\Gamma}^M dk \, \partial_k \log [\det(Q(k_x,k_y))]\right\}e^{-\frac{i}{2}[\delta \phi(M) -\delta \phi(\Gamma)]}.
 \end{align}
Similarly, the change in the Pfaffian is given by 
\begin{align}
    \pf[Q'(\bm{k}_{\rm inv})C_4 U_T]=\pf[Q(\bm{k}_{\rm inv})C_4 U_T] e^{i \delta \phi(\bm{k}_{\rm inv}) / 2},
\end{align}
where we multiply Eq.~(\ref{eq:infini_Q}) by $\det[C_4 U_T] $, take the square root of it, and use the fact that $\det[Q'(\bm{k}_{\rm inv})]$ is continuously deformed to $\det[Q(\bm{k}_{\rm inv})]$, i.e., they have the same sign of the square root. As a result, Eq.~(\ref{eq:c4_inv2}) is invariant under an infinitesimal deformation.

Now we study the $\mathbb{Z}_2$ invariant in the weak-coupling limit. We follow the procedure of Ref.~\onlinecite{Qi2010}. In the following, we assume that the Fermi surfaces are all non-degenerate and there is no superconducting nodes. Note that our discussion is generalized to the case of degenerate Fermi surfaces as we will discuss later. When the Fermi surfaces are non-degenerate and the pair potential is finite only around the Fermi surfaces, the matrix elements of $U_T \Delta^{\dagger}(k_x,k_y) $ between different bands are negligible.  
Thus, we have
\begin{align}
   A(k_x,k_y) = \mathcal{E}(k_x,k_y) - i U_T \Delta^{\dagger}(k_x,k_y) \simeq \sum_n [\epsilon_{n}(k_x,k_y) - i \delta_n(k_x,k_y)]|n,(k_x,k_y) \rangle \langle n,(k_x,k_y) |, \label{eq:app_A}
\end{align}
where 
\begin{align}
    \delta_n(k_x,k_y) \equiv \langle n,(k_x,k_y) | U_T \Delta^{\dagger}(k_x,k_y) |n,(k_x,k_y) \rangle,
\end{align}
and $ |n,(k_x,k_y) \rangle$ are the eigenstates of $\mathcal{E}(k_x,k_y) $. Since $U_T \Delta^{\dagger}(k_x,k_y)$ is a Hermitian matrix,  $\delta_n(k_x,k_y) $ takes a real value. In this approximation, $Q(k_x,k_y)$ becomes
\begin{align}
 Q(k_x,k_y) =  \sum_n e^{i \theta_n (k_x,k_y)}|n,(k_x,k_y) \rangle \langle n,(k_x,k_y) |, \label{eq:app_Q}
\end{align}
where $e^{i \theta_n (k_x,k_y)} = [\epsilon_{n}(k_x,k_y) - i \delta_n(k_x,k_y)]/|\epsilon_{n}(k_x,k_y) - i \delta_n(k_x,k_y)|$. In the weak-coupling limit, we take $\delta_n(k_x,k_y)$ to be nonzero only in the small neighborhood of the Fermi level. Hence, $e^{i \theta_n (k_x,k_y)} = \pm 1$ at momenta away from the Fermi surfaces. Since $\epsilon_{n}(k_x,k_y) < 0$ ($>0$) when it is below (above) the Fermi level, the phase $\theta_n (k_x,k_y)$ changes from $0$ to $\pm \pi$ across the Fermi level, where the sign of $\pi$ is determined from the sign of $\delta_n (k_x,k_y)$; for instance, when evaluating it along the one-dimensional line parameterized by $k$, $\theta_{n}(k)$ changes from $\pi$ ($-\pi$) to $0$ across the Fermi point $k_F$ when $\delta_n(k_F) < 0$ ($>0$) and $\partial_k \epsilon_{n}( k_F) >0$. See Fig. 1 of Ref.~\onlinecite{Qi2010}, where the sign of $\delta_n(k_F)$ is opposite to our case.  On the other hand, when $\partial_k \epsilon_{n}( k_F) <0$, $\theta_n(k)$ changes from 0 to $\pi$ ($-\pi$) for $\delta_n(k_F)<0$ ($>0$). 

We focus on the Fermi points on the $\Gamma$-M line, which are denoted by $k_{F,s}$ ($s=1,2,\cdots, M $). In the following, we assume that the Fermi points are not located at $\bm{k}_{\rm inv}$. 
In the weak-coupling limit, the change in $\theta_n (k_x,k_y)$ is characterized by a formula
\begin{align}
   \sum_n \partial_k \theta_n (k_x,k_y) =  \sum_s \pi \sgn[\partial_k\epsilon_{n}(k_x,k_y) \delta_n(k_x,k_y)] \Big|_{k \to k_{F,s}} \delta (k - k_{F,s}),
\end{align}
where $k$ is the momentum along the $\Gamma$-M line.
Since $\det[Q(k_x,k_y) ] =\exp \left[ i \sum_n \theta_n (k_x,k_y)\right] $, the exponential factor of Eq.~(\ref{eq:c4_inv2}) is calculated as
\begin{align}
     \int_{\Gamma}^M dk \, \partial_k \log [\det(Q(k_x,k_y))] &=   \int_{\Gamma}^M dk \,  i \sum_n \partial_k  \theta_n (k_x,k_y) \notag \\
     &=i \pi \sum_{s=1}^M \sgn(v_{F,s} \delta_{F,s}),
\end{align}
where we define $v_{F,s} \equiv \partial_k\epsilon_{n}(k_x,k_y)|_{k \to k_{F,s}}$ and $\delta_{F,s} \equiv \delta_n(k_x,k_y)|_{k \to k_{F,s}} $. Thus,
\begin{align}
    \exp \left\{ \frac{-1}{2} \int_{\Gamma}^M dk \, \partial_k \log [\det(Q(k_x,k_y))]\right\} &= \prod_s e^{-i \frac{\pi}{2} \sgn(v_{F,s} \delta_{F,s})} \notag \\
    &= \prod_s [-i \sgn(v_{F,s} \delta_{F,s})] \notag \\
    &= \prod_s [\sgn(\delta_{F,s})] \prod_s [-i \sgn(v_{F,s}) ]. \label{eq:int_wl}
\end{align}
When there are $m$ ($n$) Fermi points with $v_{F,s} >0$ ($<0$), the number of bands which are above the Fermi level at $\Gamma$ and below the Fermi level at $M$ is given by $n-m$. When $n(\bm{k}_{\rm inv})$ is the number of bands occupied at $\bm{k}_{\rm inv}$, $n-m = n(M) - n(\Gamma)$. Here, $n(\bm{k}_{\rm inv})$ must be an even integer due to the degeneracy of the orbital doublet. As a result, we find 
\begin{align}
    \prod_s [-i \sgn(v_{F,s}) ] = (-i)^m i^n = e^{-\frac{i}{2} \pi (m-n)} = (-1)^{[n(M) - n(\Gamma)]/2}. \label{eq:vf_wl}
\end{align}

Next, we evaluate $\pf[Q(\bm{k}_{\rm inv})C_4 U_T]$. Provided that there is no Fermi point at $\bm{k}_{\rm inv}$, we have $\Delta(\bm{k}_{\rm inv})=0$ in the weak coupling limit. That is, $A(\bm{k}_{\rm inv}) = \mathcal{E} (\bm{k}_{\rm inv})$, so that $Q(\bm{k}_{\rm inv})$ can be given by
\begin{align}
    Q(\bm{k}_{\rm inv}) = U_n^{\dagger}(\bm{k}_{\rm inv}) 
    \begin{pmatrix}
       \mathbbm{1}_{N_1} & 0 \\ 0 &  -\mathbbm{1}_{N_2}
    \end{pmatrix} 
    U_n(\bm{k}_{\rm inv}) ,
\end{align}
where $U_n (\bm{k}_{\rm inv})$ is an unitary matrix diagonalizing $\mathcal{E} (\bm{k}_{\rm inv})$ and $\mathbbm{1}_{N}$ is the $N \times N$ identity matrix. $N_1$ ($N_2$) is the number of unoccupied (occupied) bands. Then, the Pfaffian is evaluated as
\begin{align}
    \pf[Q(\bm{k}_{\rm inv})C_4 U_T] &= \pf \left[  U_n^{\dagger}(\bm{k}_{\rm inv}) 
    \begin{pmatrix}
       \mathbbm{1}_{N_1} & 0 \\ 0 &  -\mathbbm{1}_{N_2}
    \end{pmatrix} 
    U_n(\bm{k}_{\rm inv}) C_4 U_T \right] \notag \\
   & = \pf \left[  
    \begin{pmatrix}
       \mathbbm{1}_{N_1} & 0 \\ 0 &  -\mathbbm{1}_{N_2}
    \end{pmatrix} 
    U_n(\bm{k}_{\rm inv}) C_4 U_T U_n^t(\bm{k}_{\rm inv})  \right] \det[U_n^{\ast}(\bm{k}_{\rm inv}) ]. 
\end{align}
From Eq.~(\ref{eq:c4t-A}), we find $(C_4U_T) Q^t(\bm{k}_{\rm inv}) (C_4U_T)^{\dagger} = Q(\bm{k}_{\rm inv})$, which leads to the commutation relation,
\begin{align}
    \left[  U_n(\bm{k}_{\rm inv}) C_4 U_T U_n^t(\bm{k}_{\rm inv}), \begin{pmatrix}
       \mathbbm{1}_{N_1} & 0 \\ 0 &  -\mathbbm{1}_{N_2}
    \end{pmatrix}\right] =0.
\end{align}
This means that $U_n(\bm{k}_{\rm inv}) C_4 U_T U_n^t(\bm{k}_{\rm inv})$ is a block diagonal matrix. Thus, 
\begin{align}
    \pf \left[  
    \begin{pmatrix}
       \mathbbm{1}_{N_1} & 0 \\ 0 &  -\mathbbm{1}_{N_2}
    \end{pmatrix} 
    U_n(\bm{k}_{\rm inv}) C_4 U_T U_n^t(\bm{k}_{\rm inv})  \right] &= 
    \pf \left[  
    \begin{pmatrix}
       \mathbbm{1}_{N_1} & 0 \\ 0 &  i \mathbbm{1}_{N_2}
    \end{pmatrix} 
    U_n(\bm{k}_{\rm inv}) C_4 U_T U_n^t(\bm{k}_{\rm inv})
    \begin{pmatrix}
       \mathbbm{1}_{N_1} & 0 \\ 0 &  i \mathbbm{1}_{N_2}
    \end{pmatrix}
    \right] \notag \\
    &= \det \left[ \begin{pmatrix}
       \mathbbm{1}_{N_1} & 0 \\ 0 &  i \mathbbm{1}_{N_2}
    \end{pmatrix} \right]
    \pf [U_n(\bm{k}_{\rm inv}) C_4 U_T U_n^t(\bm{k}_{\rm inv}) ] \notag \\
    & = i^{N_2} \pf [C_4 U_T] \det [U_n^t(\bm{k}_{\rm inv}) ].
\end{align}
The Pfaffian results in
\begin{align}
    \pf[Q(\bm{k}_{\rm inv})C_4 U_T] &= (-1)^{N_2/2} \pf [C_4 U_T] \det [U_n^t(\bm{k}_{\rm inv}) ]\det[U_n^{\ast}(\bm{k}_{\rm inv}) ] \notag \\
    &=(-1)^{N_2/2} \pf [C_4 U_T],
\end{align}
where $N_2$ is the number of occupied band at $\bm{k}_{\rm inv}$ and takes an even integer. Therefore, we have
\begin{align}
    \frac{\pf[Q(M)C_4 U_T]}{\pf[Q(\Gamma)C_4 U_T]} = (-1)^{[n(M)-n(\Gamma)]/2} \label{eq:pfaffian_wl}
\end{align}
Substituting Eqs.~(\ref{eq:int_wl}), (\ref{eq:vf_wl}), and (\ref{eq:pfaffian_wl}) into Eq.~(\ref{eq:c4_inv2}), we obtain
\begin{align}
    (-1)^{\nu_4} = \prod_s \sgn(\delta_{F,s}). \label{eq:fs_z2c4}
\end{align}

Finally, some remarks are in order. First, the above result is generalized to the case with degenerate Fermi surfaces because we can add a symmetry-preserving perturbation which lifts the degeneracy without changing the $\mathbb{Z}_2$ invariant. Second, the Fermi surface formula for $\nu_6$ is obtained by the same procedure, where the symmetry line and $C_4U_T$ are replaced with the $\Gamma$-$K$ line and $A_6 U_T$ with $A_6 =C_6-C_6^{\dagger}$, respectively. Since $A_6 U_T = - A_6^{\dagger} U_T^t $ and $U_T =U_T^t$, one finds that $A(\bm{k}_{\rm inv}) A_6U_T$ is a skew symmetric matrix.

\begin{flushleft} 
{\bf S3. Surface energy spectra of two-dimensional models of $s_{\pm}$-wave superconductors}
\end{flushleft}

In this section, we study surface states of a model of 2D $s_{\pm}$-wave superconductors and show that a nearly gapless surface mode appears. We start from the BdG Hamiltonian (\ref{eq:bdg_Hamiv2}) with 
\begin{subequations}
 \label{eq:2d_model}
\begin{align}
\mathcal{E}(\bm{k}) = \,& t[\cos(k_x) +\cos(k_y)]- \mu + v_1 \sin(k_x) \sin(k_y) \sigma_z + v_2[\cos(k_x)-\cos(k_y)] \sigma_x, \\
\Delta(\bm{k}) =\, & \Delta_0 + \Delta_1 [\cos(k_x)+\cos(k_y)], 
\end{align}
\end{subequations}
where $\sigma_i$ describes the Pauli matrices of the orbital doublets, $\mu$ is the chemical potential, $t$ the hopping amplitude, $v_1$ and $v_2$ the orbital dependent hopping amplitudes, and $\Delta_0$ and $\Delta_1$ the pair potentials. We first study edge states from analytical calculation. We transform Eq.~(\ref{eq:2d_model}) to an effective low-energy Hamiltonian with the open boundary condition in the $x$ direction ($x \ge 0$), which is given by approximating $\cos(k)$ and $\sin(k)$ in the leading order of $k$ and replacing $k_x$ with $-i\partial_x$. Equation~(\ref{eq:2d_model}) is reduced to
\begin{align}
    H(-\partial_x ,k_y) \simeq \left(\tilde{\mu}-\frac{t}{2} \left( k_y^2-\partial_x^2\right)\right) \tau_z + v_1 k_y (-i \partial_x)\sigma_z \tau_z + \frac{v_2}{2} \left( \partial_x^2 + k_y^2 \right) \sigma_x \tau_z+ \left( \tilde{\Delta}_0 - \frac{\Delta_1}{2} (k_y^2 - \partial_x^2) \right)\tau_x , \label{eq:loweff_hami}
\end{align}
where $\tilde{\mu} \equiv -\mu + 2t$, $\tilde{\Delta}_0 \equiv \Delta_0+2\Delta_1$, and $\tau_i$ are the Pauli matrices in the Nambu space. Equation~(\ref{eq:2d_model}) preserves the TR symmetry ($U_T = 1 $) and $C_4$ symmetry ($C_4=i \sigma_y$). Note that the inversion symmetry $(I=\mathbbm{1})$ is also preserved in the 2D model. First, we consider the situation without the $v_2$ term. In this case, Eq.~(\ref{eq:loweff_hami}) has accidental chiral symmetry $\{\sigma_z \tau_y, H(\bm{k})\}=0$, which leads to zero-energy edge states protected by the chiral symmetry. To see this, we solve the differential equation $H(-\partial_x ,k_y)\bm{\Psi} (x,k_y)=0$ with $\bm{\Psi} (x,k_y) = (u_1(x,k_y),u_2(x,k_y),i u_1 (x,k_y),-i u_2(x,k_y))^t$. Here, we have chosen $\bm{\Psi} (x,k_y) $ such that $\sigma_z \tau_y\bm{\Psi}(x,k_y) = \bm{\Psi}(x,k_y)$.  Then, Eq.~(\ref{eq:loweff_hami}) is decomposed into two equations: 
\begin{subequations}
\begin{align}
    &\left[-v_1 k_y \partial_x  +(\tilde{\Delta}_0-i \tilde{\mu})-\frac{1}{2}(\Delta_1-it)(k_y^2-\partial_x^2)\right]u_1(x) =0, \\
    &\left[-v_1 k_y \partial_x +(\tilde{\Delta}_0+i \tilde{\mu})-\frac{1}{2}(\Delta_1+it)(k_y^2-\partial_x^2)\right]u_2(x) =0.
\end{align}
\end{subequations}
The equations in terms of $u_1$ and $u_2$ are decoupled, so that $u_1$ and $u_2$ give independent zero energy solutions. Provided that $u_1(x) \propto e^{\lambda_1 x} $ and  $u_2(x) \propto  e^{\lambda_2 x}$ (we omit the $k_y$ dependence), we obtain
\begin{subequations}
 \label{eq:diff_eq}
\begin{align}
    &-v_1 k_y \lambda_1 +(\tilde{\Delta}_0-i \tilde{\mu})-\frac{1}{2}(\Delta_1-it)(k_y^2-\lambda_1^2) =0, \label{eq:diff_eqa} \\
    &-v_1 k_y \lambda_2 +(\tilde{\Delta}_0+i \tilde{\mu})-\frac{1}{2}(\Delta_1+it)(k_y^2-\lambda_2^2) =0. \label{eq:diff_eqb}
\end{align}
\end{subequations}
Solving Eq.~(\ref{eq:diff_eqb}), $\lambda_2$ is given by
\begin{align}
   \lambda_{2,\pm} &= \frac{v_1 k_y}{ 2 \Delta_1'} \pm \frac{\sqrt{(v_1k_y)^2 - 4 \Delta_1' (\Delta_0'-\Delta_1' k_y^2)}}{2 \Delta_1'} \notag \\
   &\equiv \kappa_1 \pm \kappa_2 \label{eq:lambda2}
\end{align}
where  $\Delta_0' \equiv \tilde{\Delta}_0 + i \tilde{\mu}$ and $\Delta_1' \equiv (\Delta_1 + i t)/2$ and  we introduce $\kappa_1, \kappa_2$ to simplify the notation. Similarly, $\lambda_1$  is given by $\lambda_{1,\pm} = \lambda_{2,\pm}^{\ast} = \kappa_1^{\ast} \pm \kappa_2^{\ast}$. Using $\kappa_1$ and $\kappa_2$, the wave functions with zero energy are described by 
\begin{subequations}
\begin{align}
    &\bm{\Psi}_1 (x) = (u_1(x),0,i u_1 (x),0)^t  \ \ \text{with } u_1(x) = e^{\kappa_1^{\ast} x} (A_1 e^{\kappa_2^{\ast} x} + B_1 e^{-\kappa_2^{\ast} x}),\\
    &\bm{\Psi}_2 (x) = (0,u_2(x),0, -i u_2 (x))^t  \ \ \text{with } u_2(x) = e^{\kappa_1 x} (A_2 e^{\kappa_2 x} + B_2 e^{-\kappa_2 x}),
\end{align}
\end{subequations}
where $\bm{\Psi}_1 (x)$ and $\bm{\Psi}_2 (x)$ are orthogonal to each \red{other} and $A_1,B_1,A_2,B_2$ are arbitrary coefficients determined from the boundary condition and the normalization condition. We impose the boundary condition as $u_1(0)=u_2(0)=0$ and $u_1(\infty) = u_2(\infty) =0$, which lead to 
\begin{subequations}
\begin{align}
    &A_1 = -B_1, \ \ A_2 = -B_2, \label{eq:bcond1} \\
    &\Re[\kappa_1]+\Re[\kappa_2] < 0, \ \ \Re[\kappa_1]-\Re[\kappa_2] < 0, \label{eq:bcond2}
\end{align}
\end{subequations}
where $\Re[\kappa]$ is the real part of $\kappa$. 
Thus,
\begin{subequations}
\begin{align}
    &u_1(x) = 2A_1 e^{\kappa_1^{\ast} x} \sinh(\kappa_2^{\ast} x), \\
    &u_2(x) = 2A_2 e^{\kappa_1 x} \sinh(\kappa_2 x).
\end{align}
\end{subequations}
In addition, the normalization condition is imposed as $\int_0^{\infty} \bm{\Psi}^{\dagger}_i(x) \bm{\Psi}_i(x) dx=\int_0^{\infty} 2 |u_i(x)|^2 dx =1$ ($i=1,2$), which yields
\begin{equation}
   |C|^2 \equiv |A_1|^2 = |A_2|^2 =\frac{1}{2\Re[\kappa_1] |\kappa_2|^2} (\Re[\kappa_2]^2 - \Re[\kappa_1]^2) (\Re[\kappa_1]^2 + \Im[\kappa_2]^2),
\end{equation}
where $\Im[\kappa]$ is the imaginary part of $\kappa$. $\Re[\kappa_1]^2 > \Re[\kappa_2]^2$ follows from Eq.~(\ref{eq:bcond2}), so we demand $\Re[\kappa_1] <0$ in order for $|C|^2 $ to be positive. Namely, the surface zero energy states appear when $\sgn(v_1 \Delta_1 k_y) <0$. On the other hand, when considering the case that $\bm{\Psi} (x) $ satisfies $\sigma_z \tau_y \bm{\Psi}(x) = -\bm{\Psi}(x)$, $\kappa_1, \kappa_2$ in Eq.~(\ref{eq:lambda2}) changes to $-\kappa_1^{\ast}, \kappa_2^{\ast}$. Accordingly, the surface zero energy states appear when  $\sgn(v_1 \Delta_1 k_y) >0$.

We now take into account the $v_2$ term as a perturbation, which is evaluated within the first order perturbation as
\begin{align}
    &[H'(k_y)]_{ij} = \frac{v_2}{2}\int_0^{\infty} \bm{\Psi}^{\dagger}_i(x) [ (k_y^2 + \partial_x^2 ) \sigma_x \tau_z ] \bm{\Psi}_j(x), \notag \\
   &\Longrightarrow  \ \ H'(k_y)=\begin{pmatrix}
        0 & v(k_y)  \\ v^{\ast}(k_y) & 0 
    \end{pmatrix},
\end{align}
with
\begin{align}
    v(k_y)  \equiv v_2 |C|^2 \frac{\kappa_2^2 }{\kappa_1 (\kappa_2^2 - \kappa_1^2 )} (k_y^2 - \kappa_1^2 + \kappa_2^2).
\end{align}
Thus, the eigenvalues of $H'(k_y)$ are given by 
\begin{align}
    E(k_y) &=  \pm |v(k_y)| \\
    &= \pm |C|^2 \left| \frac{v_2 \kappa_2^2 }{ \kappa_1 (\kappa_2^2 - \kappa_1^2 )}\right|  \left| k_y^2 - \kappa_1^2 + \kappa_2^2\right|, \label{eq:perturb_energy}
\end{align}
where 
\begin{align}
    -\kappa_1^2 + \kappa_2^2 = k_y^2- \frac{\Delta_0'}{\Delta_1'}  = k_y^2 - 2 \frac{\tilde{\Delta}_0+ i\tilde{\mu}}{\Delta_1+it} .
\end{align}
 The last term of Eq.~(\ref{eq:perturb_energy}) is calculated as
\begin{align}
    \left| k_y^2 - \kappa_1^2 + \kappa_2^2\right| &= \left| 2 k_y^2 - 2 \frac{\tilde{\Delta}_0+ i\tilde{\mu}}{\Delta_1+it}\right| \notag \\
    &= \sqrt{\left(2 k_y^2 - 2 \frac{\tilde{\Delta}_0\Delta_1 + \tilde{\mu}t}{\Delta_1^2 +t^2}\right)^2+4\left( \frac{\tilde{\Delta}_0t - \Delta_1 \tilde{\mu}}{\Delta_1^2 +t^2}\right)^2},
\end{align}
which leads to the position of nearly gepless points $k_{y0}$ as
\begin{align}
     k_{y0} = \pm \sqrt{\frac{\tilde{\Delta}_0\Delta_1 + \tilde{\mu}t}{\Delta_1^2 +t^2}}, \label{eq:ky0}
\end{align}
where the sign of $k_{y0}$ depends on the eigenvalue of $\sigma_z \tau_y\bm{\Psi}(x)$. For instance, when $\sgn(v_1 \Delta_1) >0$, $\bm{\Psi}(x)$ with $\sigma_z \tau_y =+1(-1)$ leads to $k_{y0}<0$ $(>0)$.
Equation~(\ref{eq:ky0}) means that $k_{y0}$ is independent of $v_1$ and $v_2$. The mini gap at $k_{y0}$ is also calculated as 
\begin{align}
    2|E(k_{y0})| = 4 |C|^2  \left| \frac{v_2 \kappa_2^2 }{ \kappa_1 (\kappa_2^2 - \kappa_1^2 )}\right| \left| \frac{\tilde{\Delta}_0t - \Delta_1 \tilde{\mu}}{\Delta_1^2 +t^2}\right|. \label{eq:gap}
\end{align}
Thus, when $\tilde{\Delta}_0/\Delta_1 = \tilde{\mu}/t$, the gap closes accidentally. Away from this value, a gap arises at the edge states.

\begin{figure}[tbp]
  \begin{center}
    \includegraphics[width=\linewidth]{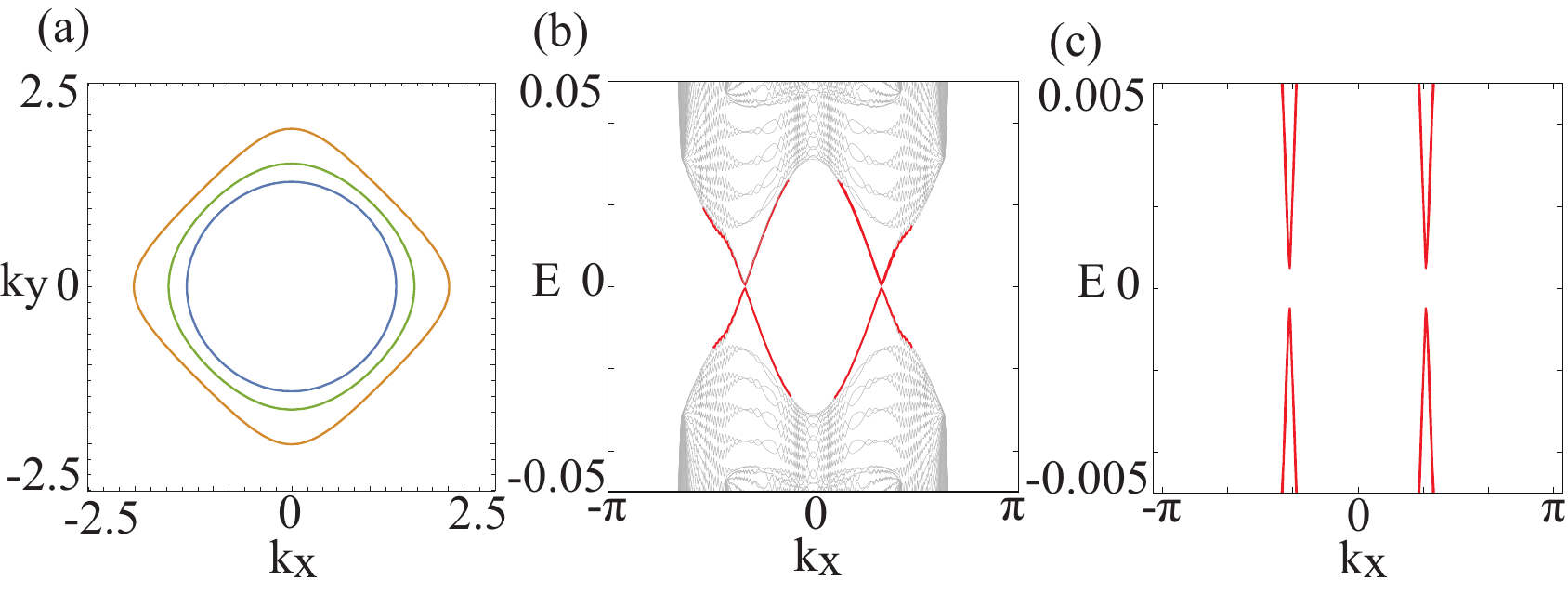}
    \caption{
    (Color online) (a) The inner (blue) and outer lines (yellow) represent the Fermi surfaces and the middle line (green) describes the zeros of the pair potential in the 2D tight-binding model (\ref{eq:2d_model}), whose parameters are chosen to be $(t,v_1,v_2,\mu)=(1,0.3,0.3,1)$. (b) The (01) surface energy spectra of Eq.~(\ref{eq:2d_model}) with $(\Delta_0,\Delta_1) =(0.1,-0.12)$, where the surface state is indicated by the thick red line. (c) The zoom of the surface state in (b). Note that a tiny gap opens at the surface state because of $C_4$ symmetry breaking.  }
    \label{fig:surface}
  \end{center}
\end{figure}

We have numerically calculated the surface energy spectrum of 2D $s_{\pm}$-wave superconductors (\ref{eq:2d_model}). Figure~\ref{fig:surface} (a) shows the Fermi surface of the normal-state Hamiltonian and the zeros of the pair potential, where we choose parameters such that two Fermi surfaces encircle the $\Gamma$ point and the sign of the pair potential changes between the two Fermi surfaces. Thus, the model corresponds to $\nu_4=1$ from Eq.~(7) in the main text. Figure~\ref{fig:surface} (b) and (c) show the surface energy spectra of the (01) edge. There are surface states with a tiny gap as shown in Fig.~\ref{fig:surface} (c), where the tiny gap arises due to breaking of $C_4$ symmetry at the boundary. Substituting $(t,\mu,v_1,v_2,\Delta_0,\Delta_1)=(1,1,0.3,0.3,0.1,-0.12)$ into Eq.~(\ref{eq:gap}), we obtain $k_{y0} \sim 1.0012$ and $2|E(k_{y0})| \sim 0.0014$, which are consistent with the numerical resutls.

\begin{flushleft} 
{\bf S4. Relationship between the $\mathbb{Z}_2$ invariant and the three-dimensional winding number}
\end{flushleft}

In this section, we study the relationship between the $\mathbb{Z}_2$ invariant $\tilde{\nu}_i$ and the three-dimensional (3D) winding number $w_{\rm 3d}$ and show that their relation is described by Eqs.~(6) and (11) in the main text. To this end, we consider the BdG Hamiltonian (\ref{eq:bdg_Hamiv2}) in 3D k-space, which satisfies the following symmetries:
\begin{align}
     & C H(\bm{k}) C^{-1} = -H(-\bm{k}), \ \ C= i\tau_y K, \label{eq:phsv3} \\
     &T H(\bm{k}) T^{-1} = H(-\bm{k}), \ \ T= \begin{pmatrix} U_T & 0 \\ 0 &U_T^{\ast}  \end{pmatrix}_\tau K, \label{eq:trsv3} \\
      &\tilde{C}_4 H(\bm{k}) \tilde{C}_4^{-1} = H(R_n \bm{k}), \ \ \tilde{C}_n= \begin{pmatrix} C_n & 0 \\ 0 &C_n^{\ast}  \end{pmatrix}_\tau, \label{eq:cns} 
\end{align}
where $\bm{k}=(k_x,k_y,k_z)$, $U_TU_T^{\ast} =(C_n)^n= \mathbbm{1}$, and 
\begin{align}
R_n \left( \begin{array}{@{\,} c @{\,}}
     k_x \\ k_y \\ k_z  
\end{array} \right)
=
\begin{pmatrix}
    \cos (2\pi/n) & -\sin (2\pi/n) & 0 \\
    \sin (2\pi/n) & \cos (2\pi/n) & 0 \\
    0 & 0 & 1
\end{pmatrix}
\left( \begin{array}{@{\,} c @{\,}}
     k_x \\ k_y \\ k_z  
\end{array} \right).
\end{align}
As discussed in Sec. S2, the BdG Hamiltonian can be continuously deformed to
\begin{align}
    VH(\bm{k})V^{\dagger} =   
     \begin{pmatrix}
        0 & A(\bm{k}) \\ A^{\dagger}(\bm{k}) & 0      
    \end{pmatrix}_{\tau}
    \sim
    \begin{pmatrix}
        0 & Q(\bm{k}) \\ Q^{\dagger}(\bm{k}) & 0      
    \end{pmatrix}_{\tau},
\end{align}
where  $Q^{\dagger}(\bm{k})Q(\bm{k}) = \mathbbm{1}$.
Using $Q(\bm{k})$, the 3D winding number is defined by
\begin{align}
    w_{\rm 3d} = \frac{1}{24 \pi^2} \int_{\rm BZ} d^3k \sum_{\alpha, \beta, \gamma} \epsilon^{\alpha \beta \gamma} \tr\left\{
    \left[Q^{\dagger}(\bm{k}) \partial_{k_\alpha} Q(\bm{k})\right] \!
    \left[Q^{\dagger}(\bm{k}) \partial_{k_\beta} Q(\bm{k})\right] \!
    \left[Q^{\dagger}(\bm{k}) \partial_{k_\gamma} Q(\bm{k})\right] \right\},
\end{align}
 where $\alpha, \beta, \gamma=x,y,z$ and $\epsilon^{\alpha \beta \gamma}$ is the Levi-Civita symbol. The winding number $w_{\rm 3d}$ takes even-integer values in class CI, but $w_{\rm 3d}=0$ in conventional $s$-wave superconductors~\cite{SatoFujimoto16}.
 It has been proved in the weak-coupling limit that $w_{3d}$ is reduced to~\cite{Qi2010}
\begin{align}
    w_{\rm 3d} = \frac{1}{2} \sum_s \sgn(\delta_s) C_s,\label{eq:w3d_fs}
\end{align}
where $s$ is summed over all disconnected Fermi surfaces and $\sgn(\delta_s)$ describes the sign of $\delta_n(\bm{k})$ on the $s$th Fermi surface (FS$_s$).  The Chern number $C_s$ of the FS$_s$ is defined by
\begin{align}
    C_s = \frac{1}{2\pi} \int_{{\rm FS}_s} \sum_{\alpha \beta} d \Omega^{\alpha \beta}[\partial_{k_\alpha} A_{s \beta}(\bm{k}) - \partial_{k_\beta} A_{s \alpha}(\bm{k})], 
\end{align}
with $A_{s \alpha} = -i \langle s \bm{k} | \partial_{k_{\alpha}} | s \bm{k} \rangle$ the Berry connection from the Bloch function $| s \bm{k} \rangle$ of the band forming the FS$_s$, and $d \Omega^{\alpha \beta}$ the surface element two forms of the Fermi surface. 

In the following, we address $C_4$ symmetric systems with orbital doublets, i.e., the $C_4$ rotation operator satisfies $(C_4T)^2 = -\mathbbm{1}$ at the $C_4T$ invariant momenta. In the presence of $T$ and $C_4$, there appears the double degeneracy of the orbital doublets at the $C_4T$ invariant momenta $\bm{k}_{\rm inv}$: $\Gamma=(0,0,0)$, $M=(\pi,\pi,0)$, $Z=(0,0,\pi)$, and $A=(\pi,\pi,\pi)$. Other accidental band degeneracies can be removed by symmetry-preserving perturbations. Thus, provided that the Fermi surface does not intersect with the $C_4T$ invariant momenta, every Fermi surface is non-degenerate (we neglect the spin degeneracy). Moreover, we can deform the Fermi surfaces into a set of spheres without changing the Chern numbers if the Fermi surfaces respect the $C_4$ rotation and time-reversal (TR) symmetries. 
After the deformation, there are two types of the spherical Fermi surfaces: ones enclosing a $C_4T$ invariant momentum (FS$_{C_4T}$) and the others enclosing no $C_4T$ invariant momenta (FS$_{0}$), respectively.

The Chern number of the Fermi surface FS$_s$ $\in$ FS$_{C_4T}$ is evaluated, as in Ref.~\cite{Alpin2023}, from the $C_4$ rotation eigenvalues $\exp[\frac{i}{2} \pi \lambda_4 (\bm{k})]$,  
\begin{align}
 C_s = \lambda_4(\bm{k}_{{\rm inv},s} + k_F \hat{\bm{z}}) - \lambda_4(\bm{k}_{{\rm inv},s}- k_F \hat{\bm{z}}) \mod 4,
\end{align}
where $\hat{\bm{z}}$ is the unit vector along the $z$ axis ($C_4$ rotation axis), $\bm{k}_{{\rm inv},s}$ is the $C_4T$ invariant momentum enclosed by the FS$_s$, and $k_F$ is a radius of the spherical Fermi surface. Since $\lambda_4 \in \{\pm 1, \pm 3\}$ for orbital doublets and the TR symmetry imposes $\lambda_4(\bm{k}_{{\rm inv},s}- k_F \hat{\bm{z}}) = - \lambda_4(\bm{k}_{{\rm inv},s}+ k_F \hat{\bm{z}})$, we find
\begin{align}
 C_s = \pm 2 \mod 4. \label{eq:cs_c4t}
\end{align}
On the other hand, the Chern number for FS$_s$ $\in$ FS$_{0}$ is evaluated as
\begin{align}
 \sum_{{\rm FS}_s \in {\rm FS}_0}C_s = 0 \mod 4.
 \label{eq:cs_arb}
\end{align}
due to the TR and $C_4$ rotation symmetry. 
We note that, for a Fermi surface enclosing a TR-invariant momentum, say, X=$(\pi,0,0)$, the TR symmetry $T^2=\mathbbm{1}$ imposes $C_s = 0 \mod 2$~\cite{Qi2010}, and similarly for its $C_4$-related Fermi surface enclosing Y=$(0,\pi,0)$, thereby satisfying Eq.~(\ref{eq:cs_arb}). 

Now $w_{\rm 3d}$ can be related to $\tilde{\nu}_4$ mod 4 as
\begin{align}
    (-1)^{w_{\rm 3d}/2} &= (-1)^{\frac{1}{2} \sum_s \sgn(\delta_s) C_s/2} \notag \\
    &=\prod_s [i \sgn(\delta_s)]^{C_s/2} \notag \\
    &=\prod_s [\sgn(\delta_s)]^{C_s/2}, \label{eq:w3d_mod4}
\end{align}
where we have used the sum rule $\sum_s C_s = 0$ in the last line~\cite{Qi2010}. 
Combining Eq.~(\ref{eq:w3d_mod4}) with Eqs.~(\ref{eq:cs_c4t}) and (\ref{eq:cs_arb}), we obtain
\begin{align}
    (-1)^{w_{\rm 3d}/2} &=\prod_{{\rm FS}_s \in {\rm FS}_{C_4T}} \sgn(\delta_s) \notag \\
    &=(-1)^{\nu_4(\pi) +\nu_4(0)} \notag \\ 
    &=(-1)^{\nu_4(\pi) -\nu_4(0)}  \notag \\
    &=(-1)^{\tilde{\nu}_4 }, \label{eq:fs_wc4}
\end{align}
where we have used Eq.~(\ref{eq:fs_z2c4}) in the second line.

In a similar manner, we try to relate $w_{\rm 3d}$ to $\tilde{\nu}_6$ for $C_6$ symmetric systems, in which the double degeneracy of orbital doublets arises at $C_6T$ invariant momenta: $\Gamma=(0,0,0)$, K=$(4 \pi/3,0,0)$, K'=$(2\pi/3,2\pi/\sqrt3,0)$ A=$(0,0,\pi)$, H=$(4 \pi/3,0,\pi)$, and H'=$(2\pi/3,2\pi/\sqrt3,\pi)$.
Here we assume that the electronic bands relevant to superconductivity are formed by the orbital doublets with $\ell=1$.
After the deformation of the Fermi surfaces, we label the Fermi surfaces enclosing the $C_6T$ invariant momenta as FS$_{C_6T,1}$ and FS$_{C_6T,2}$ and the other Fermi surfaces as FS$_{0}$, where FS$_{C_6T,1}$ (FS$_{C_6T,2}$) denotes the set of spherical Fermi surfaces enclosing $\Gamma$ and A (K, K', H, and H').

First, we consider FS$_s$ $\in$ FS$_{C_6T,1}$. The Fermi surfaces are invariant under the TR and $C_6$ rotation symmetries. Thus, after the same procedures, the Chern number of the Fermi surface is evaluated as
\begin{align}
 C_s &= \lambda_6(\bm{k}_{{\rm inv},s} + k_F \hat{\bm{z}}) - \lambda_6(\bm{k}_{{\rm inv},s}- k_F \hat{\bm{z}}) \mod 6 \notag \\ 
 &= 2\lambda_6(\bm{k}_{{\rm inv},s} + k_F \hat{\bm{z}}) \mod 6,  \label{eq:cs_c6}
\end{align}
 where the $C_6$ rotation eigenvalue at momentum $\bm{k}$ is defined as $\exp[\frac{i}{3}\pi \lambda_6(\bm{k})]$ and the second line is due to the TR symmetry.
 As we have assumed, the orbital doublets with $\ell=1$ take $\lambda_6 \in \{\pm 1\}$, so that Eq.~(\ref{eq:cs_c6}) is reduced to
 \begin{align}
 C_s = \begin{cases}
  2\mod 6 &
  \lambda_6(\bm{k}_{{\rm inv},s}+k_F\hat{\bm{z}}) =1,\\ 
  -2 \mod 6  &
  \lambda_6(\bm{k}_{{\rm inv},s}+k_F\hat{\bm{z}}) = -1.
       \end{cases} \label{eq:cs_c6t}
\end{align}

Second, the Fermi surfaces FS$_s$ $\in$ FS$_{C_6T,2}$ are invariant under the $C_2T$ and $C_3$ rotation symmetries. Thus, the Chern number is evaluated using the eigenvalues of the $C_3$ rotation symmetry $\exp[\frac{2i}{3} \pi \lambda_3(\bm{k})]$ such that
\begin{align}
 C_s = \lambda_3(\bm{k}_{{\rm inv},s} + k_F \hat{\bm{z}}) - \lambda_3(\bm{k}_{{\rm inv},s}- k_F \hat{\bm{z}}) \mod 3. \label{eq:cs_c3}
\end{align}
Since the $C_2T$ symmetry imposes $\lambda_3(\bm{k}_{{\rm inv},s}- k_F \hat{\bm{z}}) = - \lambda_3(\bm{k}_{{\rm inv},s}+ k_F \hat{\bm{z}})$ and the orbital doublets have $\lambda_3 \in \{\pm 2\}$, we obtain  
 \begin{align}
 C_s = \begin{cases}
  1 \mod 3  &
  \lambda_3(\bm{k}_{{\rm inv},s}+k_F\hat{\bm{z}})=2,\\ 
  -1 \mod 3  &
  \lambda_3(\bm{k}_{{\rm inv},s}+k_F\hat{\bm{z}})=-2,
       \end{cases} \label{eq:cs_c3t}
\end{align}
where $\lambda_3$ and $\lambda_6$ are related as $2\lambda_6=\lambda_3 \mod 3$.
In addition, there is another Fermi surface with the same Chern number related by the $C_2$ rotation symmetry. Thus, the total Chern number of the Fermi surfaces belonging to FS$_{C_6T,2}$ is $\pm 1 \times 2$. 

Finally, when FS$_s$ $\in$ FS$_{0}$, it is evaluated as
\begin{align}
 \sum_{{\rm FS}_s \in {\rm FS}_0}C_s =0 \mod 6,
 \label{eq:cs_arb6}
\end{align}
due to the TR and $C_6$ rotation symmetries.
As a result, $w_{\rm 3d} \mod 6$ is calculated as
\begin{align}
    (-1)^{w_{\rm 3d}/3} &= (-1)^{ \frac{1}{2} \sum_s \sgn(\delta_s) C_s/3} \notag \\
    &=\prod_s [i \sgn(\delta_s)]^{ C_s/3} \notag \\
    &=\prod_s [\sgn(\delta_s)]^{C_s/3} \notag \\
    &=\prod_{{\rm FS}_s \in {\rm FS}_{C_6T,1}\bigcup{\rm FS}_{C_6T,2}} \sgn(\delta_s)^{C_s/3 }. \label{eq:w3d_c6}
\end{align}
Since the Fermi surfaces enclosing K' and H' points are related to those enclosing K and H by $C_6$ symmetry, Eq.~(\ref{eq:w3d_c6}) is written as
\begin{align}
    (-1)^{w_{\rm 3d}/3} &=
    \prod_{s; \Gamma{\rm -K}}[\sgn(\delta_s)]^{C_s/3}
    \prod_{s; {\rm A-H}}[\sgn(\delta_s)]^{C_s/3},
    \label{eq:w3d_c6-2}
\end{align}
where the products are taken for the Fermi surfaces crossing either $\Gamma$-K line or A-H line, and
$C_s$ in Eq.~(\ref{eq:w3d_c6-2}) is either $+2$ or $-2$.
The formula cannot be further simplified due to the two possible values of $C_s=+2$ and $-2$.

\bibliography{spin_singlet_tsc}